\documentclass[%
 reprint,
superscriptaddress,
 amsmath,amssymb,
 aps,
floatfix,
]{revtex4-1}

\usepackage{graphicx}
\usepackage{dcolumn}
\usepackage{bm}
\usepackage{soul,color}
\usepackage{float}
\usepackage{multirow}
\usepackage{color}
\usepackage{verbatim}
\usepackage{amsmath}
\usepackage{amssymb}
\usepackage{graphicx}
\usepackage{esint}

\usepackage[unicode=true,
 bookmarks=true,bookmarksnumbered=false,bookmarksopen=false,
 breaklinks=false,pdfborder={0 0 1},backref=false,colorlinks=true]
 {hyperref}
\usepackage{lipsum}
\usepackage{xcolor}
\usepackage{soul}
\begin{document}

\preprint{APS/123-QED}


\title{Efficient and tunable frequency conversion using periodically poled thin-film lithium tantalate nanowaveguides}%
\author{Simin Yu}
\thanks{These authors contribute equally to this work.}
\author{Mingyue Qi}
\thanks{These authors contribute equally to this work.}
\author{Huizong Zhu}
\author{Bofu Zhao}
\author{Jingchun Qian}
\author{Qiushi Chen}
\affiliation{State Key Laboratory of Quantum Functional Materials, School of Information Science and Technology, ShanghaiTech University, Shanghai 201210, China}
\author{Juanjuan Lu}
 \email{lujj2@shanghaitech.edu.cn}

\affiliation{State Key Laboratory of Quantum Functional Materials, School of Information Science and Technology, ShanghaiTech University, Shanghai 201210, China}

\date{\today}

\begin{abstract}
 Thin-film lithium tantalate (TFLT) has recently emerged as a promising photonic platform for chip-scale nonlinear optics due to its weaker photorefraction, higher optical damage threshold, broader transparency window, and lower birefringence compared to that of thin-film lithium niobate. Here we develop an ultralow-loss lithium tantalate integrated photonic platform and report the first functional second harmonic generator based on high-fidelity poling of z-cut TFLT. As a result, quasi-phase matching (QPM) is performed between telecom (1550\,nm) and near-visible (775\,nm) wavelengths in a straight waveguide and prompts strong second-harmonic generation with a normalized efficiency of 229\,\%/W/cm$^2$. An absolute conversion efficiency of 5.5\% is achieved with a pump power of 700\,mW. Such a second-harmonic generator exhibits stable temperature tunability (-0.44\,nm/$^\circ\mathrm{C}$) which is important for applications that require precise frequency alignment such as atomic clocks and quantum frequency conversion.
\end{abstract}

\maketitle


\section{Introduction}
The second-order nonlinearity ($\chi^{(2)}$) is fundamental to many crucial nonlinear optical processes, including second-harmonic generation (SHG)~\cite{wangUltrahighefficiencyWavelengthConversion2018,luoHighlyTunableEfficient2018}, sum frequency generation~\cite{SFG}, and optical parametric oscillation~\cite{OPO,Maidment:16,lu2021}. Among these, SHG plays a particularly significant role in various applications, such as spectroscopy~\cite{spencerOpticalfrequencySynthesizerUsing2018a,spectro}, supercontinuum generation~\cite{luSupercontinuumGeneration2019,hongSupercontinuum2023}, quantum frequency conversion~\cite{QuantumFrequency,Murakami:23}, and entangled photon-pair generation~\cite{luChipintegratedVisibleTelecom2019,EntangledPhotonPair}. Compared with other popular $\chi^{(2)}$ materials like aluminum nitride~\cite{JungTang+2016+263+271,liuAluminumNitrideNanophotonics2021} and gallium arsenide~\cite{kuo2014,GaAs}, ferroelectric materials including potassium titanyl phosphate~\cite{Driscoll:86,Brown:92}, lithium niobate (LN)~\cite{Lu:19,yangSymmetricSecondharmonicGeneration2024} and lithium tantalate (LT)~\cite{liu2024,chen2025} stand out due to their large second-order nonlinear coefficients and flexibility in ferroelectric domain control. LN, in particular, has attracted considerable attention with the advent of thin-film lithium niobate (TFLN) technology, which has greatly advanced photonic integrated circuits, allowing the design of compact and high-performance optoelectronic chips~\cite{vazi2022,feng2024}. 

However, despite its popularity, TFLN has certain limitations like low optical damage threshold and strong photorefract effect, restricting its performance under high power. Strategies such as material doping, improved crystal growth, and post-processing have been explored, but they also bring new challenges~\cite{CHEN2023109753,Younesi:24}. As a result, the quest for novel thin-film ferroelectric materials has become a central focus of current research initiatives. The recent demonstration of high-quality thin-film lithium tantalate (TFLT) has established it as an excellent alternative to TFLN. TFLT exhibits a comparable refractive index ($n=2.12$) and second-order nonlinearity ($d_{33}=26$\,pm/V) to TFLN. Moreover, TFLT demonstrates an enhanced optical damage threshold (240\,mW/cm$^2$), a broader transparent window ($0.28-5.5\,\mathrm{\mu m}$), and a lower birefringence (0.004) further enhancing its potential for devices including electro-optic modulators, frequency converters, and optical switches~\cite{Wang2024,LTPR,LTapplications}. Several SHG devices based on intermodal phase-matching~\cite{modal2024} and periodically poled lithium tantalate (PPLT) on x-cut~\cite{PPLT2025} have already been developed, showcasing its promise for nonlinear photonic applications. 

In this paper, we present the development of an ultralow-loss integrated z-cut TFLT photonic platform and report the first functional second-harmonic generator based on high-fidelity poling. As a result, quasi-phase matching is realized between telecom (1550\,nm) and near-visible (775\,nm) wavelengths in a straight waveguide, yielding strong SHG with a normalized efficiency of 229\,\%/W/cm$^2$. A maximum absolute conversion efficiency of 5.5\% is realized at pump power of 700\,mW. Notably, this second harmonic generator exhibits stable temperature tunability (-0.44\,nm/$^\circ\mathrm{C}$), which is important for applications that require precise frequency alignment, such as atomic clocks and quantum frequency conversion.

\section{Device Design and Fabrication}

\begin{figure}[t]
\includegraphics[width=0.45\textwidth]{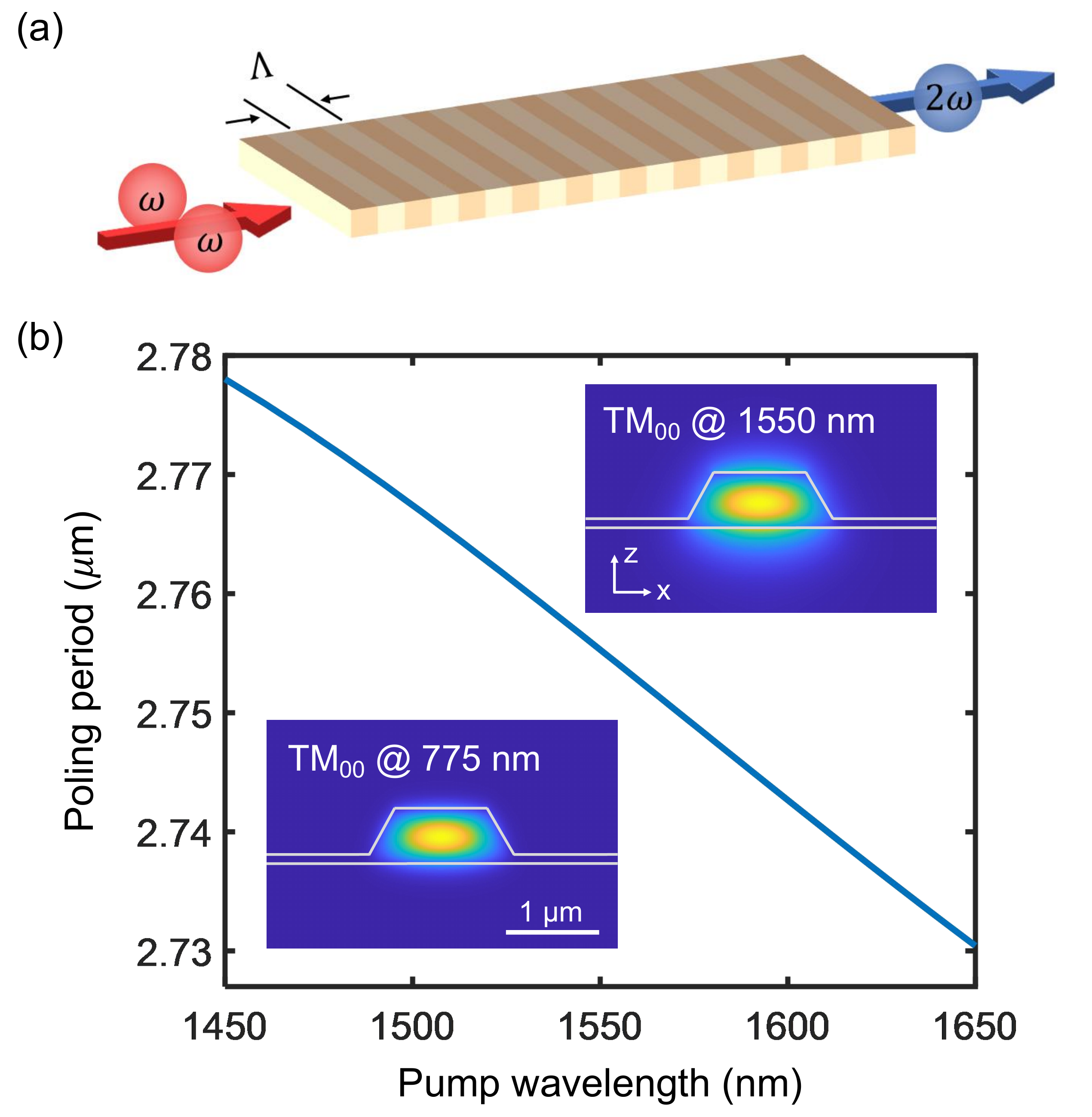}  
\caption{\label{design}(a) Schematic of SHG process in a PPLT waveguide, where the annihilation of two fundamental photons generates a second-harmonic photon. (b) Simulated poling period as a function of pump wavelength with a fixed film thickness of 600\,nm and unetched layer of 100\,nm, the refractive index is established in ref~\cite{index1996}. The insets show the simulated electric field distributions for the fundamental (top) and second harmonic (bottom) TM modes.}
\end{figure}

Figure\,\ref{design}(a) illustrates the design principle of the PPLT waveguide, where the SHG process produces a photon with twice the frequency by combining two photons of the fundamental frequency. The LT waveguide has a fixed width of 1\,$\mathrm{\mu m}$ and a thickness of 600\,nm with 100\,nm thick unetched layer. The poling period $\Lambda$ for QPM SHG at room temperature is determined by $\Lambda=\lambda_{2\omega}/(n_{2\omega}-n_{\omega})$, where $\lambda_{2\omega}$ is the second-harmonic wavelength, while $n_{\omega}$ and $n_{2\omega}$ are the effective refractive indices at the first-harmonic (FH) and second-harmonic (SH) wavelengths, respectively. To utilize the highest second-order nonlinear tensor component $d_{33}$, we simulate the poling period for the conversion of the fundamental transverse magnetic (TM$_{00}$) mode from FH to SH wavelength, as shown in Fig.\ref{design}(b). The required poling period is estimated to be $\sim2.75\,\mu m$ at a pump wavelength of 1550\,nm. The inserts show the numerically simulated optical mode profiles of TM$_{00}$ mode at both FH (1550\,nm) and SH (775\,nm) wavelengths.

\begin{figure}[htbp]
\includegraphics[width=0.45\textwidth]{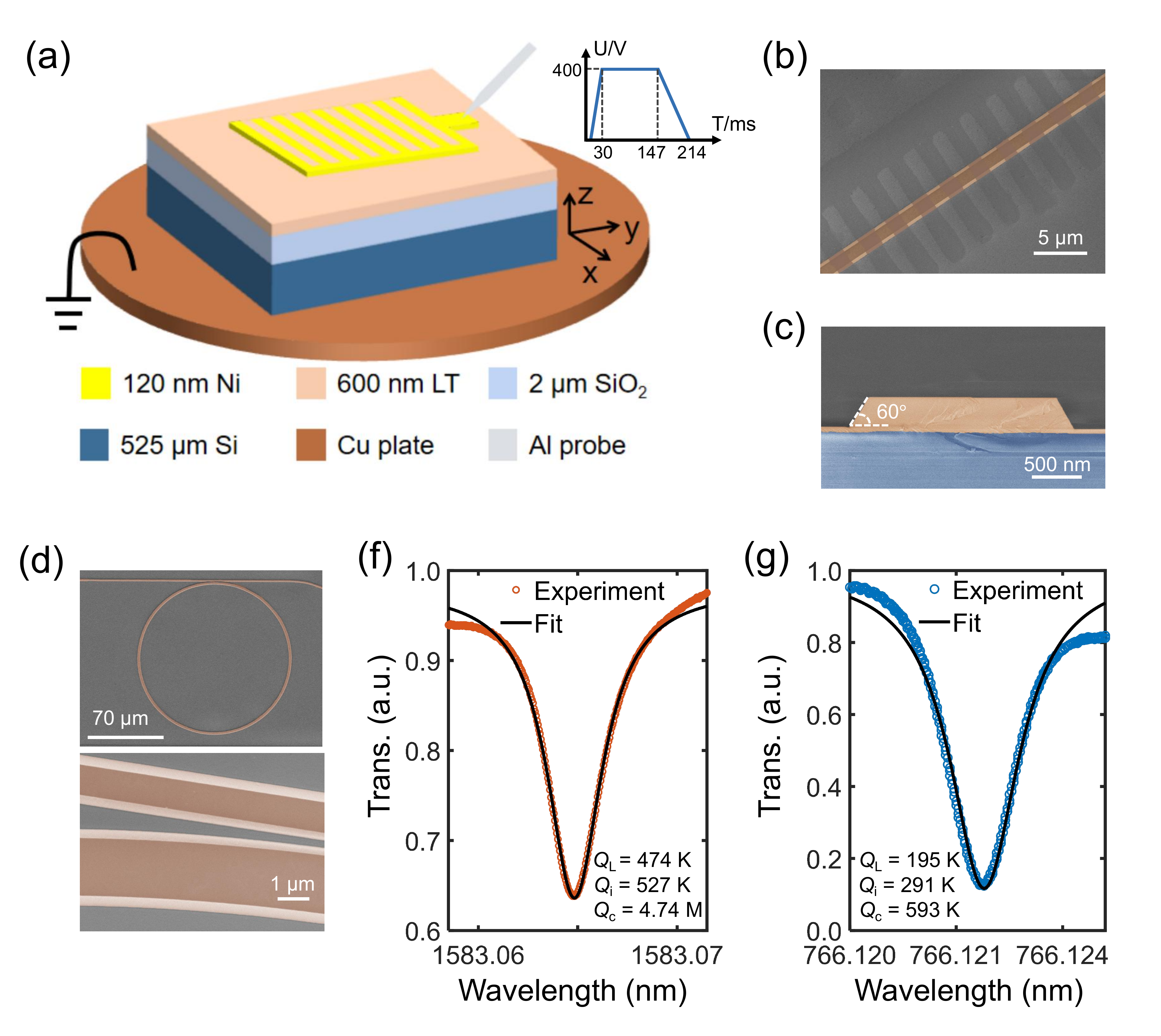}  
\caption{\label{poling}(a) Schematic diagram of the poling setup for z-cut LTOI devices. The insert shows the waveform of poling poling voltage pulse. (b-d) False-color SEM images of a PPLT waveguide (b), the cleaved waveguide facet with a sidewall angle of 60$^\circ$ (c), a microring and its sidewall in the coupling region (d), respectively. (e-f) Lorentz fit of TM mode at FH wavelength (e) and SH wavelength (f) with extracted loaded ($Q\mathrm{_L}$), intrinsic ($Q\mathrm{_i}$), and coupling ($Q\mathrm{_c}$) $Q$ values.}
\end{figure}

\begin{figure*}[htbp]
\includegraphics[width=0.9\textwidth]{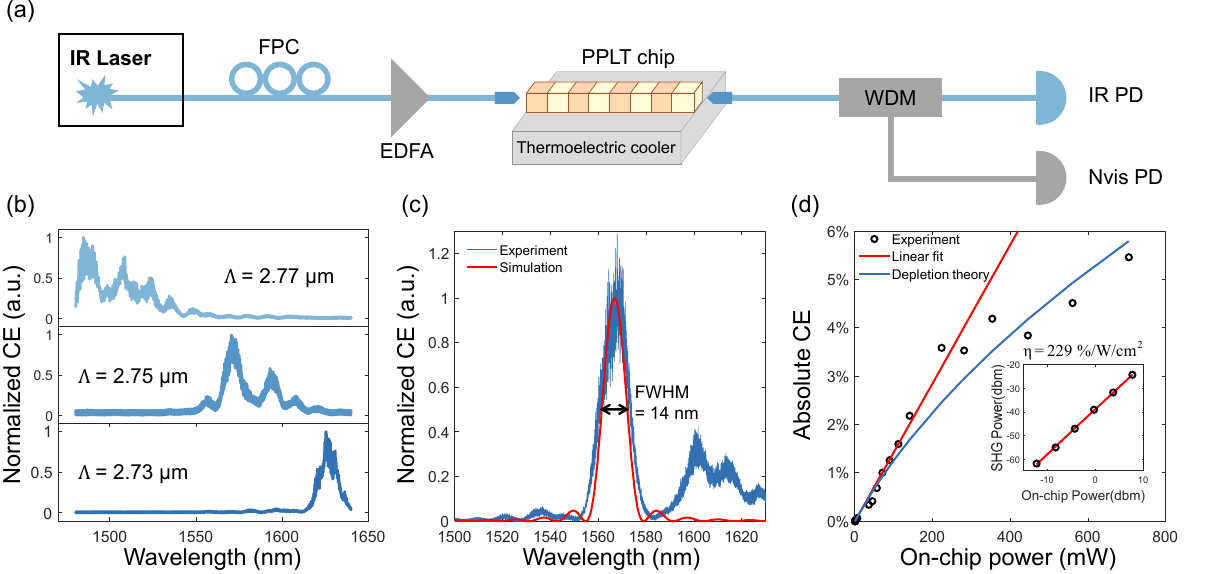}  
\caption{\label{setup}(a) Illustration of experimental setup for the characterization of the PPLT waveguides. FPC, fiber polarization controller; EDFA, erbium-doped fiber amplifier; WDM, wavelength division multiplexer; PD, photodetector. (b) SHG spectra of PPLT waveguides with varying poling periods at 25\,$^\circ\mathrm{C}$. (c) Normalized SHG efficiency versus pump wavelength, indicating a FWHM of 14\,nm and matching well with the numerical simulation. (d) Absolute conversion efficiency as a function of on-chip pump power. The inset presents the SHG-pump power relation in the non-depleted regime.}
\end{figure*}

The device fabrication commences with the patterning of waveguides, followed by the poling process. A commercial lithium niobate on insulator (LTOI) wafer (supplied by NANOLN) is utilized, consisting of a 600\,nm-thick z-cut LT thin film on a 2.0\,$\mu$m-thick silicon dioxide (SiO$_2$) layer over a silicon substrate. The bus waveguide pattern is defined using electron beam lithography (EBL) with hydrogen silsesquioxane resist and developed in 25\% TMAH for high contrast. An optimized inductively coupled plasma reactive ion etching (ICP RIE) with Ar$^+$ plasma transfers the pattern onto the LT layer. The chip is subsequently immersed in a solution of 3:1 KOH (40\%):H$_2$O$_2$ (30\%) for 3 hours at $40^\circ\mathrm{C}$ to remove the redeposition generated by dry etching. For the poling process, nickel (Ni) finger electrodes are first deposited on the LT waveguides through EBL and liftoff processes. The chip is heated to $250^\circ$ on a cuprum plate and then subjected to three 400\,V, 120\,ms pulses via a probe, as depicted in Fig.\,\ref{poling}(a). The waveform of applied voltage pulse is shown in the inset. After removing Ni, the inverted domains are clearly visible by scanning electron microscope (SEM), exhibiting a duty cycle approaching 50\%, as illustrated in Fig.\ref{poling}(b). The high poling fidelity is essential for achieving high conversion efficiency. The bus waveguide is ultimately tapered to a width of 3\,$\mu$m at both facets to improve the fiber-to-chip coupling efficiency. Figure\,\ref{poling}(c) shows the cleaved waveguide facet with a sidewall angle of 60$^\circ$. The insertion losses are calibrated to be -5.22 and -5.44\,dB/facet for the telecom and near-visible wavelengths, respectively. Figure\,\ref{poling}(d) displays the false-color SEM images of a microring (top) and its smooth sidewall (bottom) in the coupling region. The propagation loss of the TFLT waveguide is characterized by measuring the optical quality factor ($Q$) of the microring. A Lorentzian fit is respectively applied to the resonance dips of the TM mode transmission spectra around 1583\,nm and 766\,nm, as shown in Figs.\,\ref{poling}(e-f), yielding propagation loss of 0.72\,dB/cm in the telecom wavelength and 2.69\,dB/cm in the near-visible wavelength.

\section{Results and discussion}

\subsection{Second-harmonic generation}
Figure\,\ref{setup}(a) depicts the experimental setup for SHG measurement and device characterization. A tunable telecom laser (Santec TSL570) serves as the pump source, with a fiber polarization controller (FPC) ensuring that the on-chip pump light is aligned to the TM polarization. The telecom (IR) and near-visible (Nvis) outputs are separated using a wavelength division multiplexer (WDM) and subsequently measured by the corresponding photodetectors (PD). The phase-matching wavelengths of devices, measured at 25\,$^\circ\mathrm{C}$, for poling periods of 2.77, 2.75 and 2.73\,$\mathrm{\mu m}$, are found at 1485, 1571 and 1626\,nm, respectively, as shown in Fig.\,\ref{setup}(b). These experimental results align well with the simulations presented in Fig.\,\ref{design}(b), confirming both the effectiveness and precision of the poling process. Figure\,\ref{setup}(c) shows a typical $sinc^2$-like normalized conversion efficiency spectrum with a phase-matching wavelength of 1568\,nm, exhibiting a full width at half maximum (FWHM) of 14\,nm. The slight deviation between the simulated (red line) and experimental data (blue line) is possibly attributed to the non-formality of poling and film thickness. To further increase the FWHM bandwidth for applications requiring broad spectral operation, aperiodic poling design and waveguide dispersion engineering could be employed. 

The power dependence of the conversion efficiency has also been investigated through an erbium-doped fiber amplifier (EDFA) to amplify the optical power from the pump laser. The highest absolute conversion efficiency is measured to be 5.5\,\% at a pump power of 700\,mW, as shown in Fig.\,\ref{setup}(d). The experimental pump depletion behavior aligns well with the theoretical prediction\,\cite{000043}. A linear fit in the non-depleted regime suggests a quadratic dependence of SHG power on pump power (inset of Fig.\,\ref{setup}), resulting in an on-chip normalized SHG efficiency of 229\,\%/W/cm$^2$. We note that the recorded normalized efficiency is lower than the theoretical value, which is mainly attributed to the material defects (vacancy and inhomogeneity of the LT thin film) and poling imperfections (deviation of poling period and duty cycle). Further improvement could be envisioned with high-quality thin-film lithium tantalate on insulator wafers and adapted control on the poling period~\cite{10.1038/s41565-023-01525-w}. Additionally, optimizing waveguide geometry and implementing post-fabrication annealing could help mitigate propagation losses and improve phase matching.

\subsection{Thermal tunability}
\begin{figure}[htbp]
\includegraphics[width=0.45\textwidth]{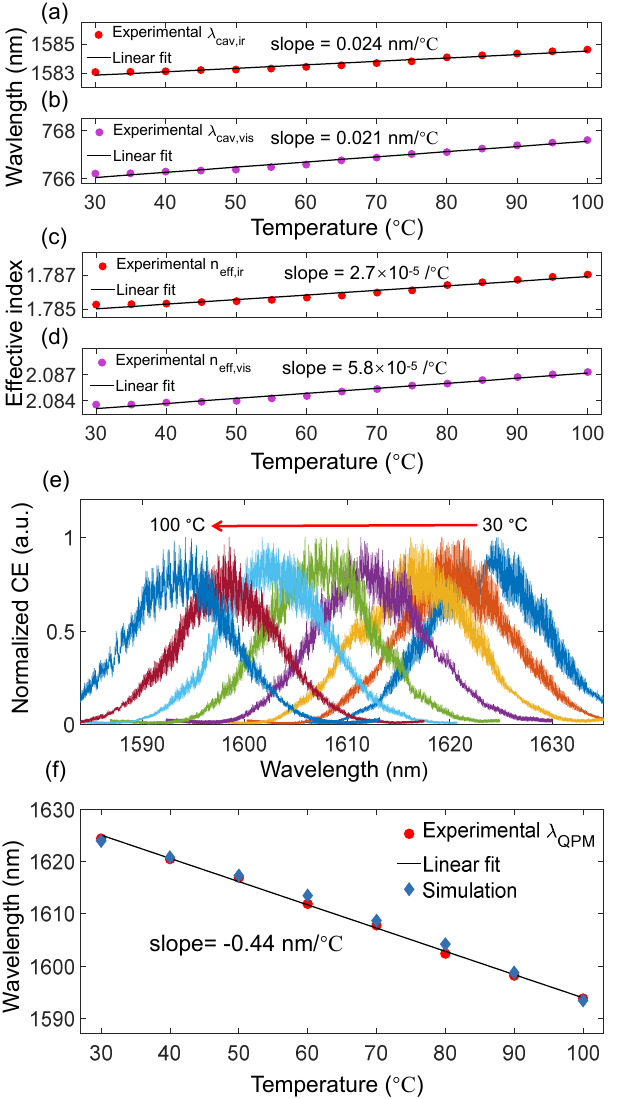}  
\caption{\label{tempdependence}Measured temperature-dependent cavity resonant wavelength at the respective telecom (a) and near-visible (b) wavelengths. (c-d) Extracted effective index versus temperature from (a-b). (e) Blue shift of the measured SHG spectra as the temperatures increased in increments of 10\,$^\circ\mathrm{C}$ from 30\,$^\circ\mathrm{C}$ to 100\,$^\circ\mathrm{C}$. (f) QPM wavelength temperature dependency corresponding to (e), with the experimental tunability fitted to be -0.44\,nm/$^\circ\mathrm{C}$.}
\end{figure}

We characterize the thermal response of the LT microring resonator. Figures \ref{tempdependence}(a) and (b) depict the temperature-dependent resonant wavelength $\lambda_{cav}$ for the modes in Figs.\,\ref{poling}(f-g), yielding thermal responses of 24\,pm/$^\circ\mathrm{C}$ at the telecom band and 21\,pm/$^\circ\mathrm{C}$ at the near-visible band, respectively. Based on the relation $2\pi Rn_{eff}(T)=m\lambda_{cav}(T)$, where $R$ is the radius of microring, $n_{eff}$ is the effective index of resonant mode, $m$ is the mode number, we extract the temperature dependence of the effective index $d[n_{eff}(T)]/dT$ for both modes [Figs.\,\ref{tempdependence}(c-d)], which indicates the efficient tunability of our microring cavity.

The thermal tunability of our QPM waveguide is further studied. Figure\,\ref{tempdependence}(e) clearly shows a blue shift of the SHG peak wavelength with increasing temperature, exhibiting a thermal tunability of -0.44\,nm/$^\circ\mathrm{C}$ as implied in Fig.\,\ref{tempdependence}(f). In addition to the thermal-optical effect of the LT material $n_{e(o)}(T)$\,\cite{index1996}, pyroelectricity in LT generates an internal electric field $E_z$ along the z-axis at elevated temperatures, therefore induces refractive index variation $\Delta n$ via the Pockels effect\,\cite{10.3390/cryst14070579,10.61011/FTT.2024.11.59332.257}:
\begin{align}
\Delta n_{e(o)}(T)&=-\frac{1}{2} n^3_{e(o)}(T) r_{33(13)} E_z(T),
\\
E_z(T)&=-\frac{p}{\varepsilon_o \varepsilon_r} (T-25^\circ\mathrm{C}).
\label{pockel}
\end{align}
Here, $r_{33}$=\,27.4\,pm/V and $r_{13}$=\,6.92\,pm/V are the electro-optic coefficients of LT; $\varepsilon_0$ and $\varepsilon_r$=\,31 are the vacuum and relative dielectric constants; $p$=\,-230\,µC/(m$^{2}$K) is the pyroelectric coefficient\,\cite{p230}. And the thermal expansion of the waveguide modulates the mode confinement and poling period, given by\,\cite{doi:10.1063/1.1657244}:
\begin{align}
\begin{aligned} 
h(T)= &\ h\left(25^{\circ} \mathrm{C}\right)\left[1+0.22 \times 10^{-5}\left(T-25^{\circ} \mathrm{C}\right)\right. \\
& \left.-5.9 \times 10^{-9}\left(T-25^{\circ} \mathrm{C}\right)^2\right],
\end{aligned}\\
\begin{aligned} 
\quad X(T)= &\ X\left(25^{\circ} \mathrm{C}\right)\left[1+1.62 \times 10^{-5}\left(T-25^{\circ} \mathrm{C}\right)\right. \\
& \left.+5.9 \times 10^{-9}\left(T-25^{\circ} \mathrm{C}\right)^2\right],
\end{aligned}
\label{expan}
\end{align}
where $h$ is the thickness of z-cut TFLT waveguide and $X$ denotes the width of waveguide $W$ and poling period $\Lambda$.

Numerical modeling that incorporates all these three effects predicts the simulated QPM wavelength versus temperature (blue diamonds) in Fig.\,\ref{tempdependence}(f), which matches well with the experimental results (red dots). This agreement demonstrates that the observed thermal tuning arises primarily from the interplay between geometric changes due to thermal expansion, pyroelectric field-induced index modulation, and conventional thermo-optic effects. The minor deviations between simulation and experiment likely stem from unaccounted factors like waveguide dimension tolerance and temperature-dependent variations in the pyroelectric coefficient\,\cite{10.1039/c9tc05222d}. Nevertheless, the remarkable consistency between theory and experiment provides strong evidence for the reliability of our device's thermal tuning characteristics and confirms the effectiveness of our design approach for thermally tunable nonlinear photonic devices.

\section{\label{sec:intro}Conclusion}
In conclusion, we have successfully demonstrated an ultralow-loss integrated lithium tantalate photonic platform and the first functional second harmonic generator based on high-fidelity poling of z-cut TFLT. We achieve SHG with a normalized efficiency of 229\,\%/W/cm$^2$, and a maximum absolute conversion efficiency of 5.5\% at a pump power of 700\,mW. The demonstrated temperature tunability of -0.44\,nm/$^\circ\mathrm{C}$ further reinforces the potential for precise frequency alignment, which is critical for many precision applications. Our work not only offers significant insights into optimizing SHG performance, but also establishes a promising foundation for future applications in fields such as quantum frequency conversion and atomic clocks. Moreover, the performance can be further optimized by reducing propagation losses and improving poling quality, paving the way for even more impressive applications of second harmonic generator on the TFLT platform.


\vspace{2 mm}
\noindent \textbf{Acknowledgements.} This work is supported by National Key R$\&$D Program of China (2024YFB2807400) and National Natural Science Foundation of China (62305214). The facilities used for device fabrication were supported by the ShanghaiTech Material Device Lab (SMDL). J. Lu acknowledges support from the State Key Laboratory of  Photonics and Communications, Shanghai Jiao Tong University, China. 

\vspace{2 mm}
\noindent \textbf{Author contributions.} S.Y. performed the device design and fabrication. M.Q. carried out the measurements and analyzed the data. M.Q. and S.Y. prepared the figures and wrote the manuscript with contributions from all authors. J.L. supervised the project.

\vspace{2 mm}
\noindent \textbf{Competing interests.} The authors declare no competing interests.

\noindent \textbf{Data availability.} The data that support the findings of this
study are available from the corresponding author upon reasonable request.

\def\bibsection{\section*{\textbf{references}}}
\bibliographystyle{myaipnum4-1}
\bibliography{References}

\end{document}